\documentclass[]{IEEEphot}

\jvol{xx}
\jnum{xx}
\jmonth{June}
\pubyear{2020}

\graphicspath{{pictures/}}

\begin{document}

\title{Investigation of the dependence of noise characteristics of
SPAD on the gate parameters in sine-wave gated single-photon
detectors}

\author{A.~V.~Losev${}^{1,2,3,4}$, V.~V.~Zavodilenko${}^{1,4}$, A.~A.~Koziy${}^{1,4}$, Y.~V.~Kurochkin${}^{1,3,4}$, A.~A.~Gorbatsevich${}^{2,5}$}

\affil{${}^1$"Qrate" LLC, St. Novaya, d. 100, Moscow region, Odintsovo, Skolkovo village, 143026, Russia. \\
${}^2$National Research University of Electronic Technology MIET, Shokin Square, 1, Zelenograd, 124498, Russia. \\
${}^3$National University of Science and Technology MISIS, Leninsky prospect, 4, Moscow, 119333, Russia. \\
${}^4$NTI Center for Quantum Communications, National University of Science and Technology MISiS, Leninsky prospekt 4, Moscow, 119049, Russia\\
${}^5$P.N. Lebedev Physical Institute of the Russian Academy of Sciences,  Leninsky prospect, 53, Moscow, 119333, Russia.}  


\maketitle

\markboth{IEEE Photonics Journal}{?}

\begin{receivedinfo}%
    This research was sponsored by the Russian Science Foundation, grant No. 17-71-20146.
\end{receivedinfo}

\begin{abstract}
In this paper, we have investigated a self-developed sine wave gated (SWG) single-photon
detector (SPD) for 1550 nm wavelength primary for quantum key distribution (QKD) usage. We have
investigated different gate parameters’ influence on the SPD’s noise characteristics. We have admitted that with an increase of gating voltage and constant value of quantum efficiency ($QE$), the dark count rate     ($DCR$) decreases. There have been made some recommendations to improve SPD’s and whole QKD device's characteristics based on these observations. There have been discovered the quick rise of the $DCR$ value with the increase of gating voltage above some certain value and this value was different for different detectors. It has been shown that universal empirical dependence compilation to connect control and operational parameters of SPD is a non-trivial task.
\end{abstract}

\begin{IEEEkeywords}
single photon detector, single photon avalanche diode,  dark count rate, quantum efficiency.
\end{IEEEkeywords}

\section{Introduction}
\qquad SPD’s have many applications, such as QKD \cite{Kiktenko_2018}, 3D imaging systems (LIDAR)  \cite{Yu:17}, optical time domain reflectometry (OTDR)  \cite{OTD_article}, fluorescence microscopy \cite{Lee_2016}, biomarker tomography \cite{8740927}, astronomical telescopy  \cite{10.1117/12.2265785} etc. SPD’s structure is not universal, it’s designed for a specific application and operating parameters. There are several kinds of devices can be used as a single photon sensor, for example, avalanche photodiode (APD) \cite{7747499}, single photon avalanche diode (SPAD) \cite{7283534}, negative feedback avalanche diode (NFAD) \cite{NFAD_article_2}, semiconductor matrices \cite{10.1117/12.2540769}, superconducting nanowires  \cite{Li:16}.  

In this paper, we have used an SPD, based on InGaAs/InP SPAD, as an investigation object. This SPD is designed for use in compact (standard server rack compatible) QKD device  \cite{Kiktenko_2017}, This device works on 1550 nm optical wavelength. APD structure is not optimized for receive optical pulses with average energy less than one photon per pulse, so the choice of SPAD for this problem is optimal. Because of an integrated resistor, using NFAD increases the speed of the electrical circuit, but these diodes have high noise characteristics \cite{NFAD_article} due to many problems while producing. Superconducting nanowire single-photon detector (SNSPD) are large and expensive, which makes it difficult to create a compact QKD device.

SPAD control electronics  \cite{Stipcevic, Jong} has a set of important for effective diode’s work parameters, such as DC bias voltage, AC bias voltage (gating signal), the shape of gating signal, and dead time. It’s quite important but a laborious task to find the optimal shape of gating signal to make $QE$ high enough  ($10$ -- $20$ \%) and $DCR$ low enough  ($<100$ Hz). In this paper, we explore the influence of gating parameters to $QE$ and $DCR$ to make some recommendations about how to tune an SPD to increase the efficiency of a whole QKD device.

There are SPD’s functional scheme and the influence analysis of the gating parameters on $QE$, $DCR$ and afterpulsing probability (AP) in section 2. There are SPD’s measuring stand scheme and methodology for characterizing SPD's parameters in section 3. There are experimental data analysis and general recommendations for improving SPD’s performance in section 4. There is summary in section 5.

\section{Electrical circuit and the influence of gating parameters on the physical processes in SPAD}
\qquad The development of electrical control circuitry of SPAD to provide certain operations with diode is an  important task when design SPD. In this paper, the electrical circuit, which provides a periodical crossing of the SPAD to Geiger mode, passive avalanche quenching with active reset and dead time is used  \cite{10.1117/12.685808}. We consider the dead time as the time interval when SPD is not sensitive to optical radiation.

\begin{figure}[h]
\centering
\includegraphics[width=0.8\linewidth]{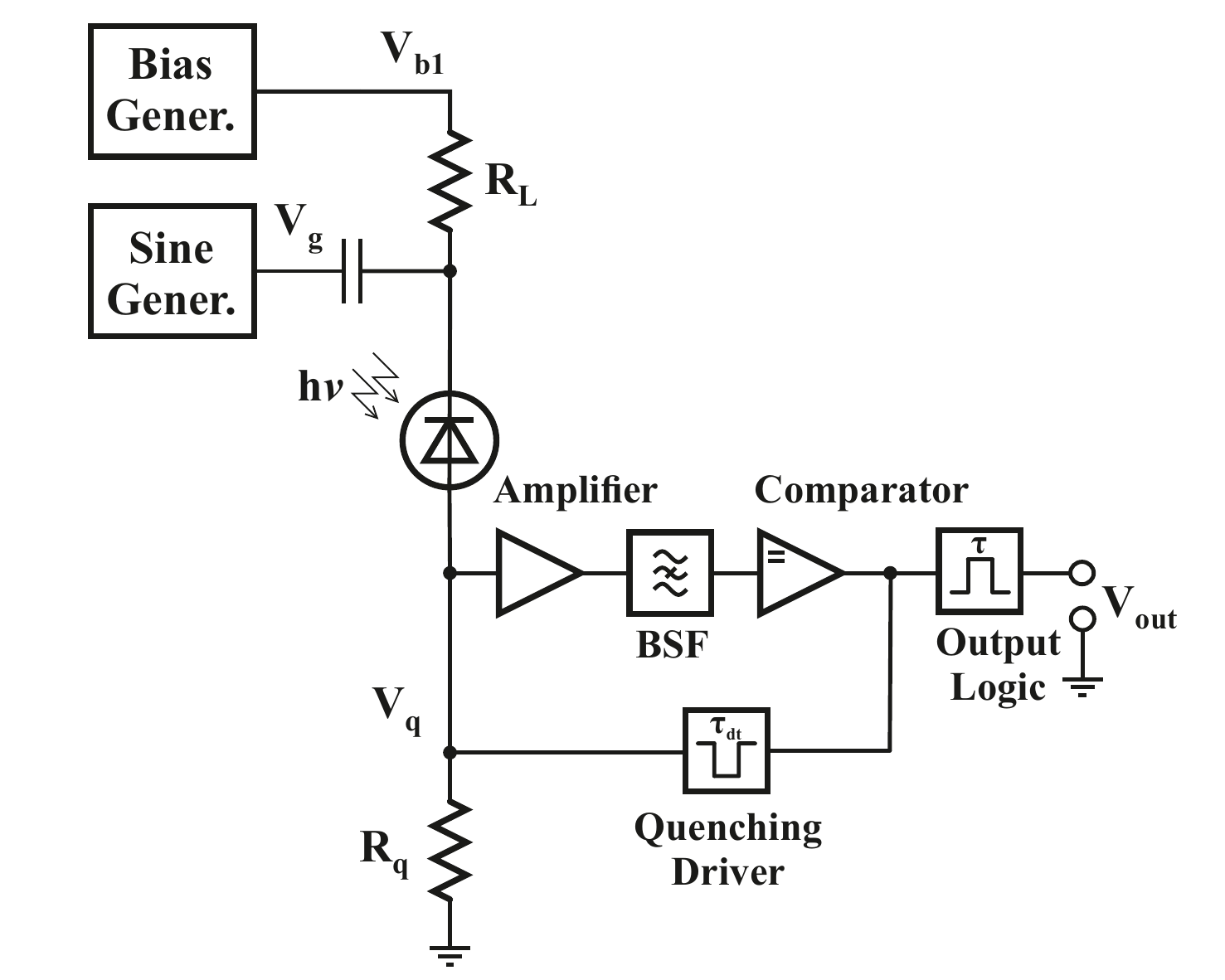}
\caption{Functional circuit of SPD. In the circuit: Bias Gener. -- DC bias generator; Sine Gener. --  alternating sinusoidal bias voltage (gate) generator with a frequency of $312.5$ MHz; BSF -- the barrage filters to eliminate the influence of gates on the processing of avalanche signals.}
\label{fig:el_sch}
\end{figure}

The functional diagram of the developed SPD is shown in Figure   \ref{fig:el_sch}. SPAD is reverse biased with the direct bias voltage  $V_{b1}$ and alternate bias voltage  $V_g$ from the cathode side. The alternate bias voltage (gates) is a $312.5$ MHz  frequency sine electrical wave. Terminating resistor  $R_L$ and quenching resistor  $R_{q}$ are used to quench the avalanche and reset the SPAD after a breakdown. There are amplifiers and suppression filters to eliminate the gating signal influence on the avalanche signal processing on the anode side of the SPAD. Comparator converts analog avalanche signal to a certain logic level digital one. This signal arrives at the output logic block and the quenching driver simultaneously. Output logic block makes a signal with certain amplitude and duration that goes to the receiver. Quenching driver makes a dead time pulse. It decreases the direct bias voltage by a fixed amount  $V_q$ within a $\tau_{dt}$ time.

The most important and determining SPAD’s operating mode output parameters of the presented electrical circuit are the parameters of the AC and DC bias voltages, the shape of the gating signal, and the dead time. The maintaining temperature of the SPAD has also particular importance. These parameters must be selected for each specific used SPAD since its performance can vary significantly from diode to diode even in the same batch \cite{compstudy}. This approach allows us to achieve the best performance of SPD.

\begin{figure}[h]
\centering
\includegraphics[width=1\linewidth]{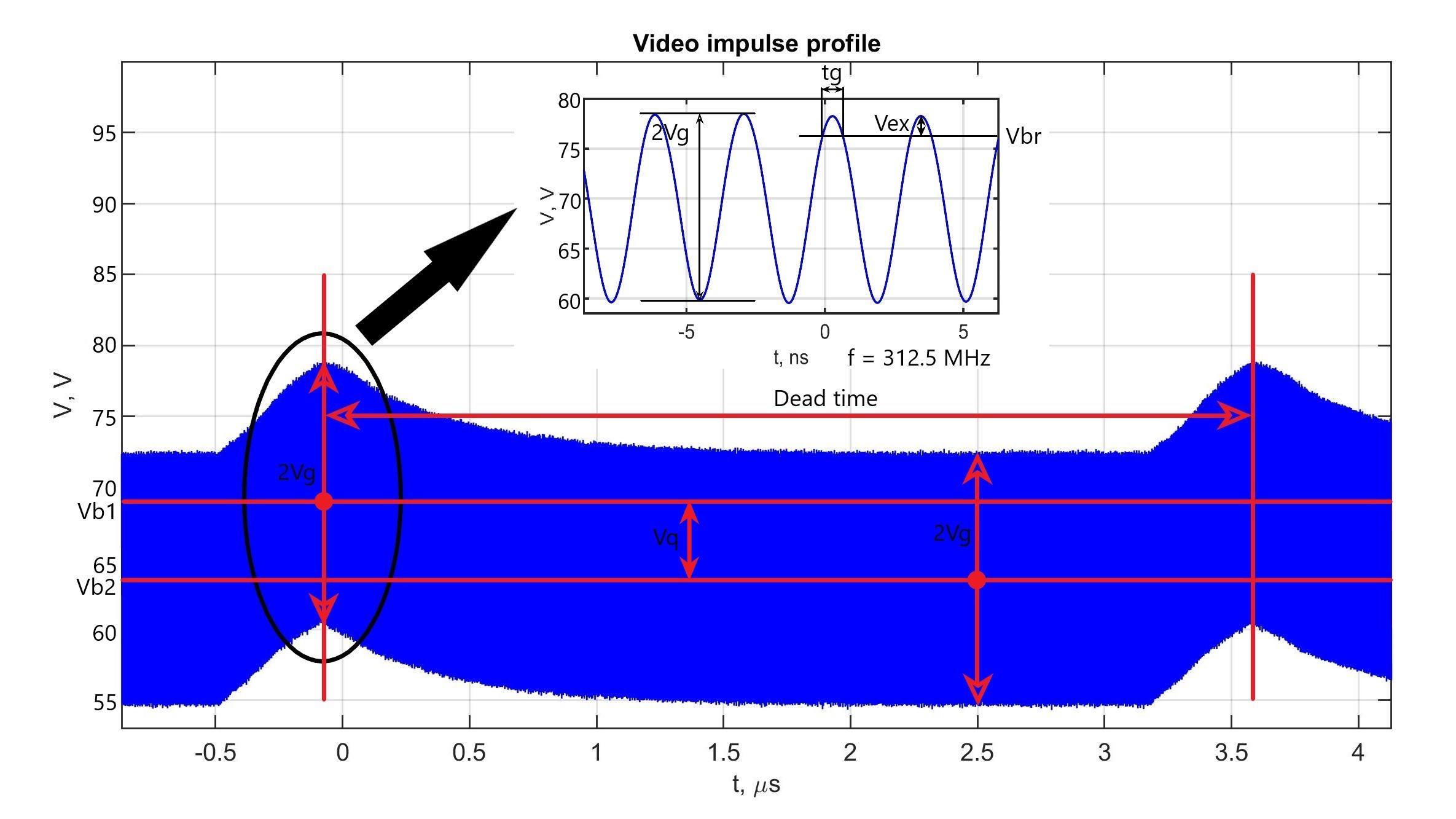}
\caption{Bias voltage on SPAD.}
\label{fig:pic_time}
\end{figure}

\qquad AC bias voltage is a $312.5$ MHz frequency sinusoidal signal with amplitude  $V_g$ peak-peak. It is applied to SPAD’s cathode in addition to the DC bias voltage $V_{b1(b2)}$. Gating voltage has to exceed SPAD’s breakdown voltage  $V_{br}$ to make single photon detection possible on a value  $V_{ex}$ within a $t_g$ time (Geiger mode time). Direct bias voltage $V_{b1}$ falls to $V_{b2}$ after detection to exclude the probability of following avalanche breakdown. The dead time pulse shape is made as shown in Figure  \ref{fig:pic_time} to reduce the effect of transient processes on single photon detection. The time when the bias voltage does not exceed the breakdown voltage of the SPAD is called dead time $\tau_{dt}$. The operation mode of SPAD described above is called gated mode with dead time.

It is Geiger mode when  $V_a> V_{br}$. Caught in SPAD’s multiplication region free carriers generate a positive feedback avalanche process. Free carriers in SPAD’s heterostructure can appear due to photogeneration effect, direct and indirect tunneling, relaxation of traps, and other less probable effects, like Poole-Frenkel effect and photon-assisted generation through defects \cite{6504460}. Photogeneration and thermal generation prevail in the absorption region because of the low energy gap of the InGaAs material, as is shown in \cite{6504460, 6247522}.  The tunneling generation effect is determined with the energy gap and the value of electric field in material. It takes place in absorption and multiplication regions.

Traps for minority charge carriers, electrons, prevail in InGaAs/InP SPAD’s heterostructure. Relaxation of these traps can trigger avalanches. This effect is called the afterpulsing. It imposes some serious restrictions on key distribution speed in QKD systems. We don’t focus on the afterpulses in this work. We have minimized contribution to SPD’s noise by afterpulses due to the fixed value of dead time equal to  $\approx 3.5 \ \mu s$, so we have not considered afterpulsing further.

Such SPD parameters like direct bias voltage $V_{b1}$, gate signal shape, and amplitude $V_g$ define maximum SPAD’s excess voltage $V_{ex}$ and diode’s time in Geiger mode. These values can be estimated from equations \ref{eq:1} and \ref{eq:2}.

\begin{equation} \label{eq:1}
V_{ex} = V_{b1} + V_g - V_{br},
\end{equation}

\begin{equation} \label{eq:2}
t_g = \frac{1}{\omega} \left(\pi - 2 \arcsin \left(\frac{V_{br} - V_{b1}}{V_g}\right)\right),
\end{equation}

where $\omega$ -- is circular frequency of sinusoidal gating signal.

\begin{figure}[h]
\begin{minipage}[h]{0.47\linewidth}
\center{\includegraphics[width=1\linewidth]{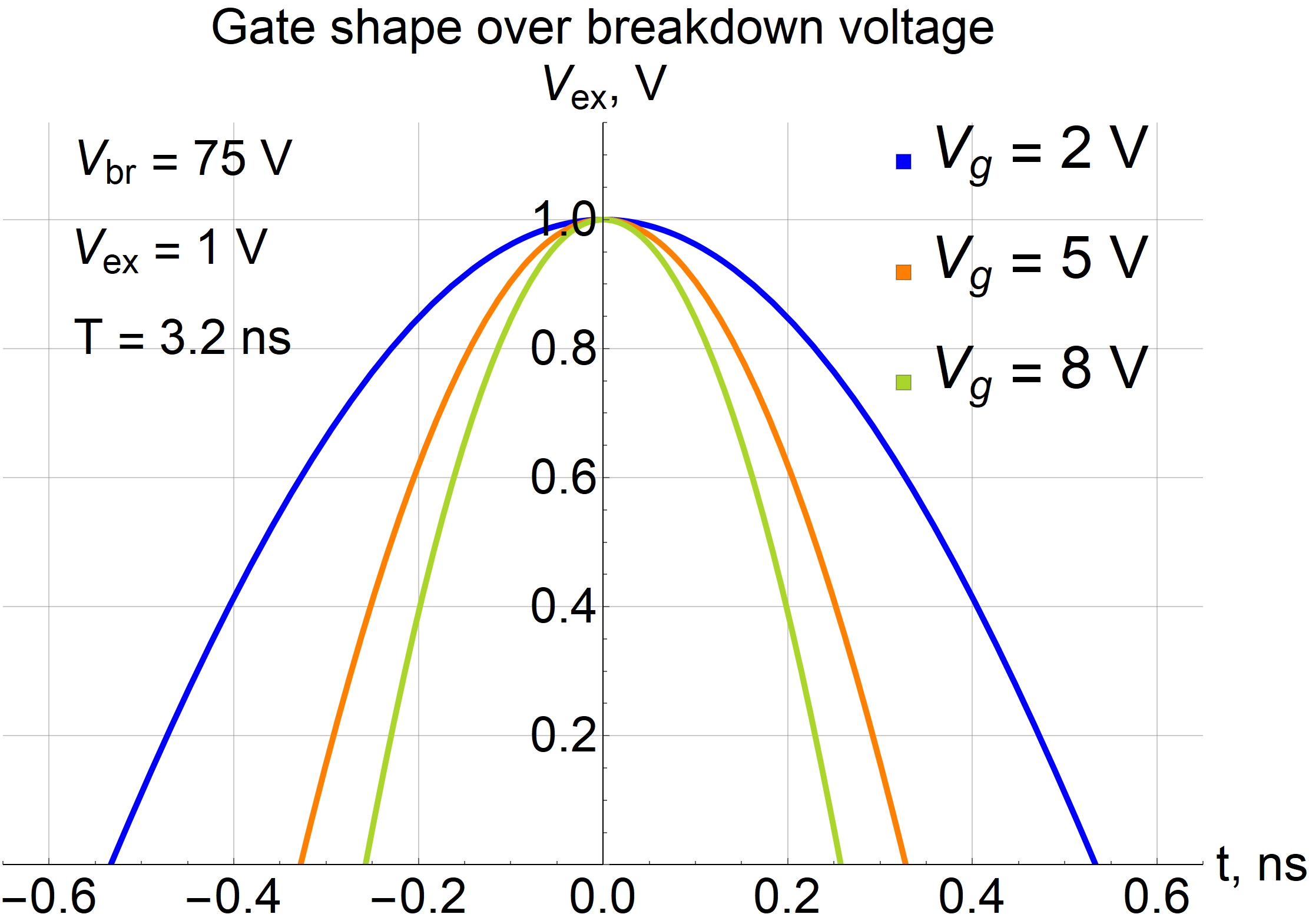}} a) \\
\end{minipage}
\hfill
\begin{minipage}[h]{0.47\linewidth}
\center{\includegraphics[width=1\linewidth]{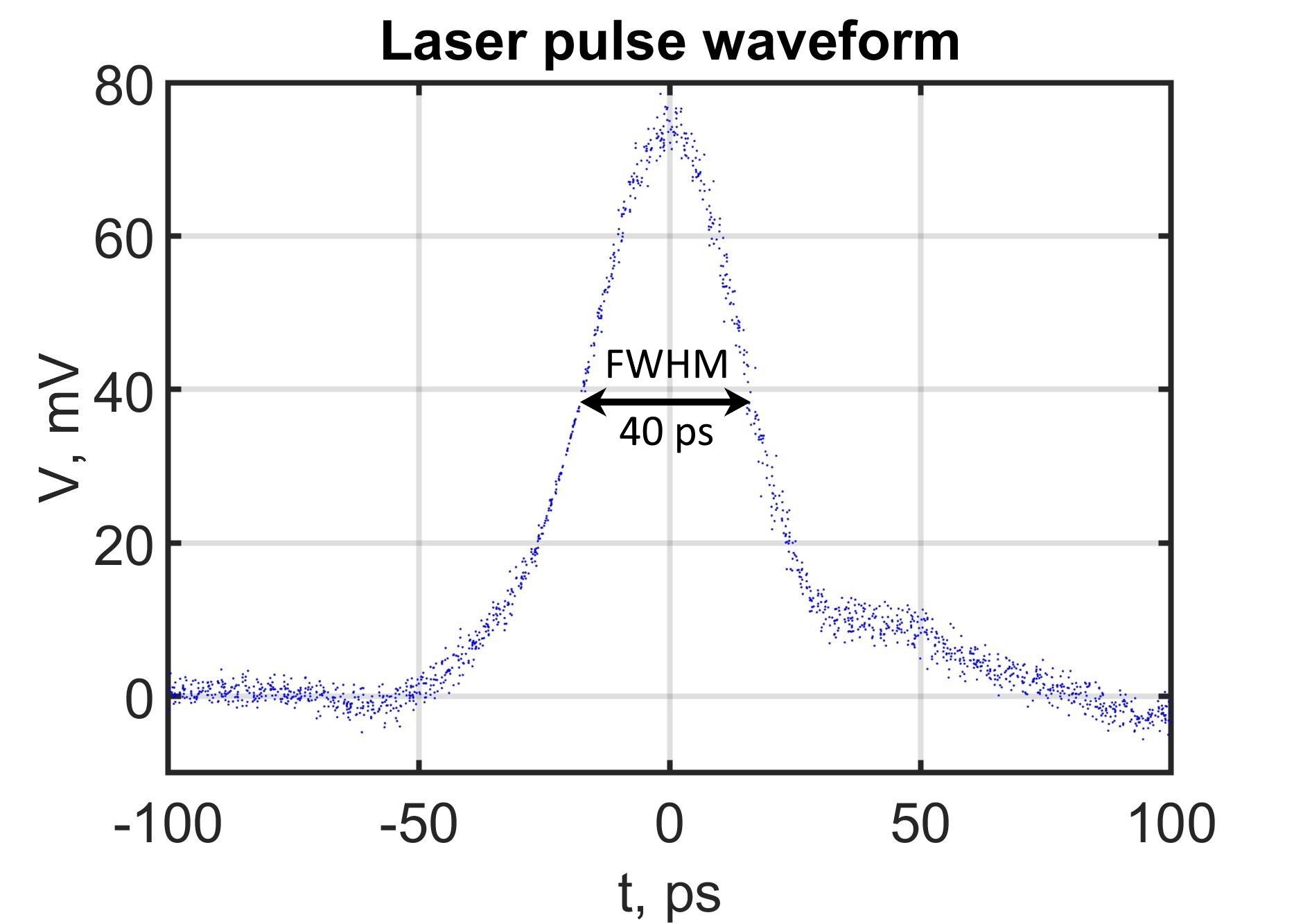}} \\b)
\end{minipage}
\vfill
\begin{minipage}[h]{0.47\linewidth}
\center{\includegraphics[width=1\linewidth]{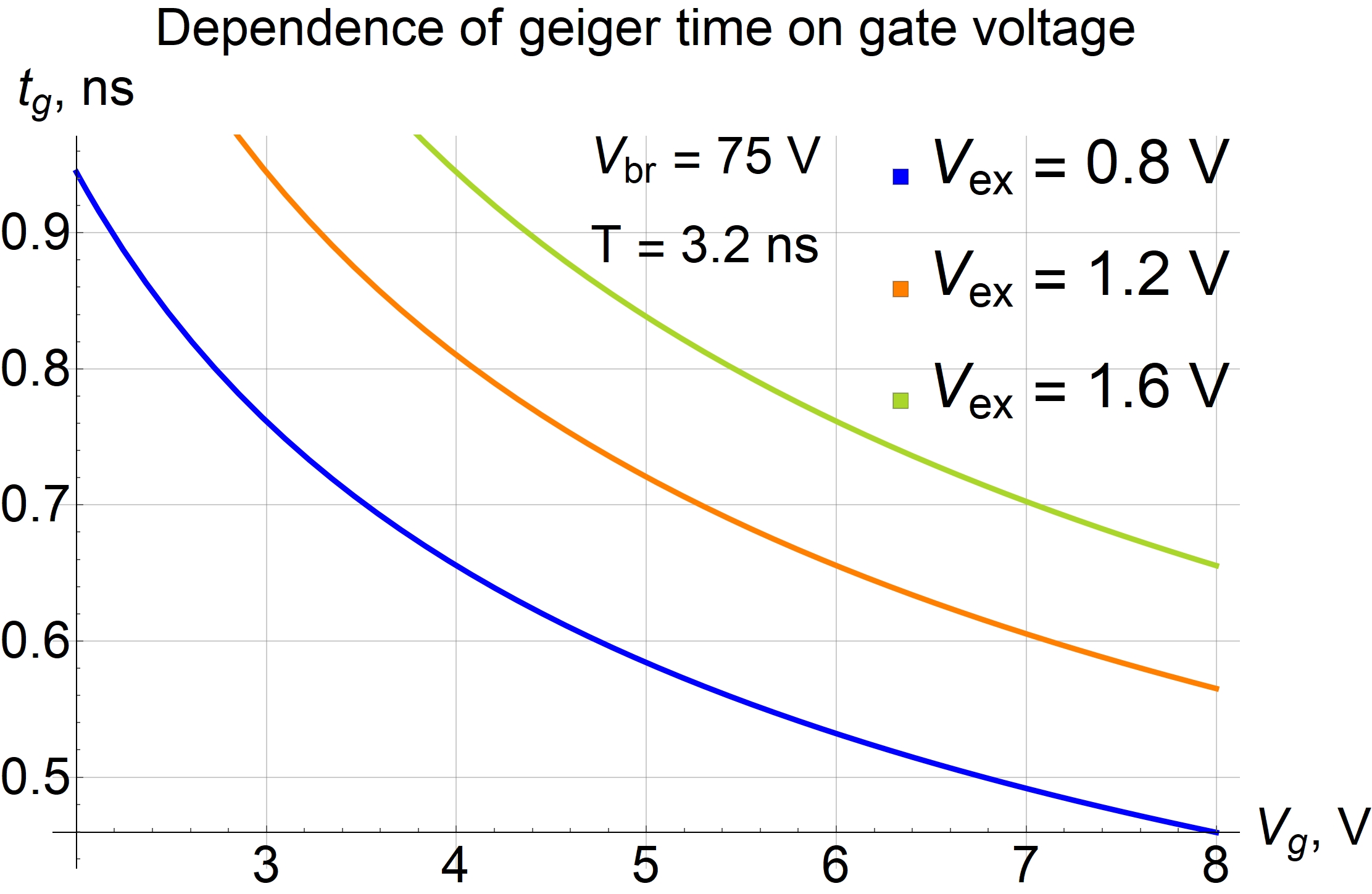}} c) \\
\end{minipage}
\hfill
\begin{minipage}[h]{0.47\linewidth}
\center{\includegraphics[width=1\linewidth]{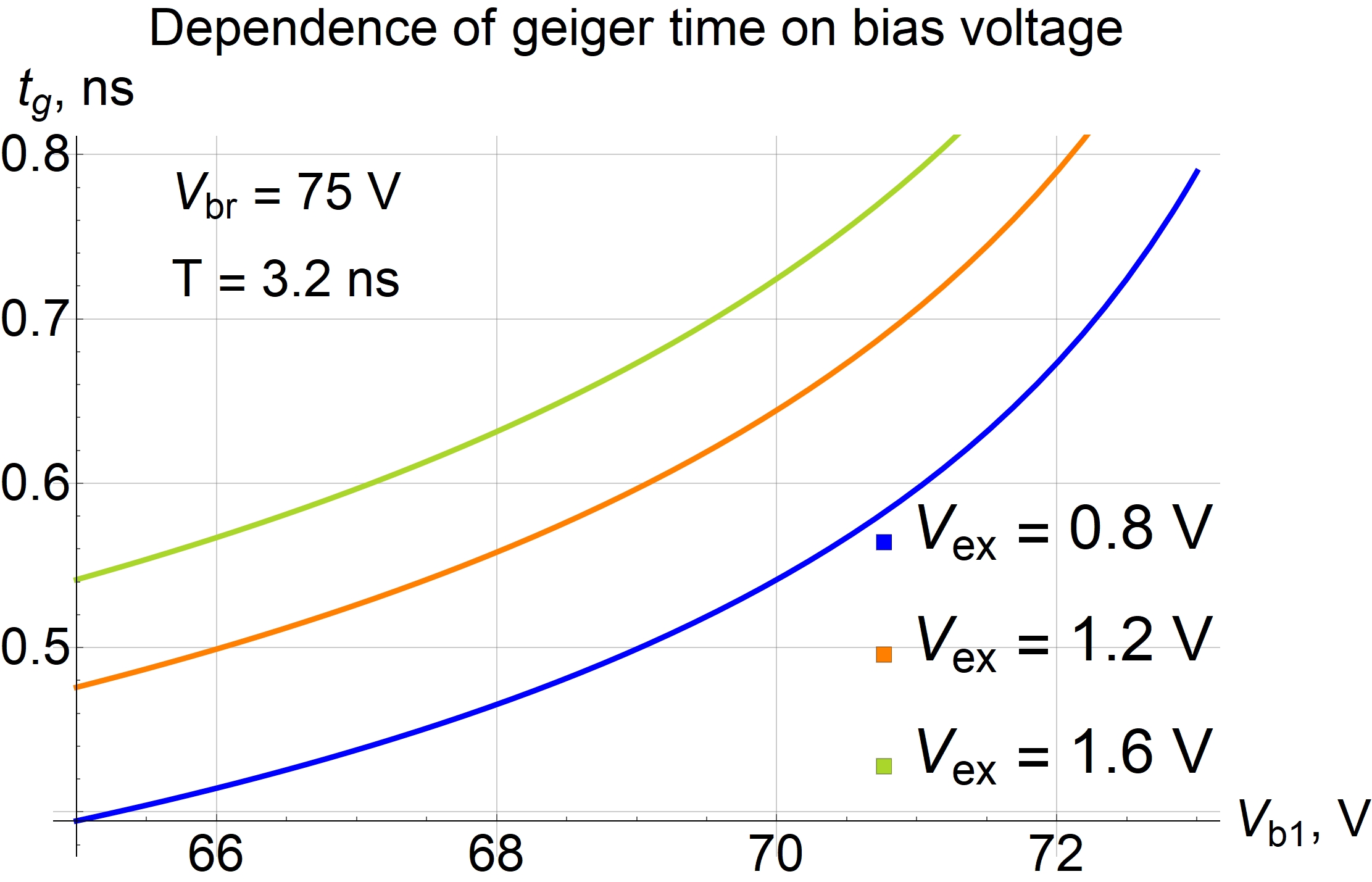}} d) \\
\end{minipage}
\caption{a) The shape of excess voltage  $V_{ex}$ for various values of gate amplitude:  $V_{g} = \{2, 5, 8\}$  V and excess voltage  $V_{ex} = 1$ V; b) the waveform of the laser pulse obtained at the measuring stand (see Fig. \ref{fig:meas_sch}), with FWHM equal to  $40$  ps; c) the dependence of the geiger mode time $t_g$ on the gate amplitude  $V_g$ for various values of the excess voltage $V_{ex} = \{0.8, 1.2, 1.6\}$  V; d) the dependence of the geiger time $t_g$ on the direct bias voltage  $V_{b1}$ for various values of the excess voltage  $V_{ex} = \{0.8, 1.2, 1.6\}$  V.  General parameters:  $V_{br} = 70$ V, $T = 3.2$ ns.}
\label{fig:gate1}
\end{figure}

There are analytical graphs for the shape of SPAD’s excess voltage $V_{ex}$ under different gate amplitude shown in Figure \ref{fig:gate1}a. There is a laser pulse waveform from the measuring stand (see Fig. \ref{fig:meas_sch}) shown in Figure \ref{fig:gate1}b. There are $t_g$ versus $V_g$ under different $V_{ex}$ diagrams shown in Figure \ref{fig:gate1}c. Also, we have placed $t_g$ versus $V_{b1}$ under different excess voltage values $V_{ex}$ diagrams in Figure \ref{fig:gate1}d.

The rise of excess voltage increases the probability of an avalanche excitation with free charge carriers in the multiplication region. Increase of Geiger mode time $t_g$ causes an increase of dark count probability per gate, including afterpulsing, and $QE$. Timing jitter of the laser pulse and the SPAD determine the distribution parameters of a random variable – the moment of avalanche generation. An increase in gating time increases the probability of photogeneration charge carrier to excite the avalanche process in SPAD. So, a variation of gating parameters leads to a change of $QE$ and $DCR$, which ratio we have to optimize to achieve an effective operating mode of SPD.

\section{SPD’s performance parameters measurement}
\begin{figure}[h]
\centering
\includegraphics[width=0.8\linewidth]{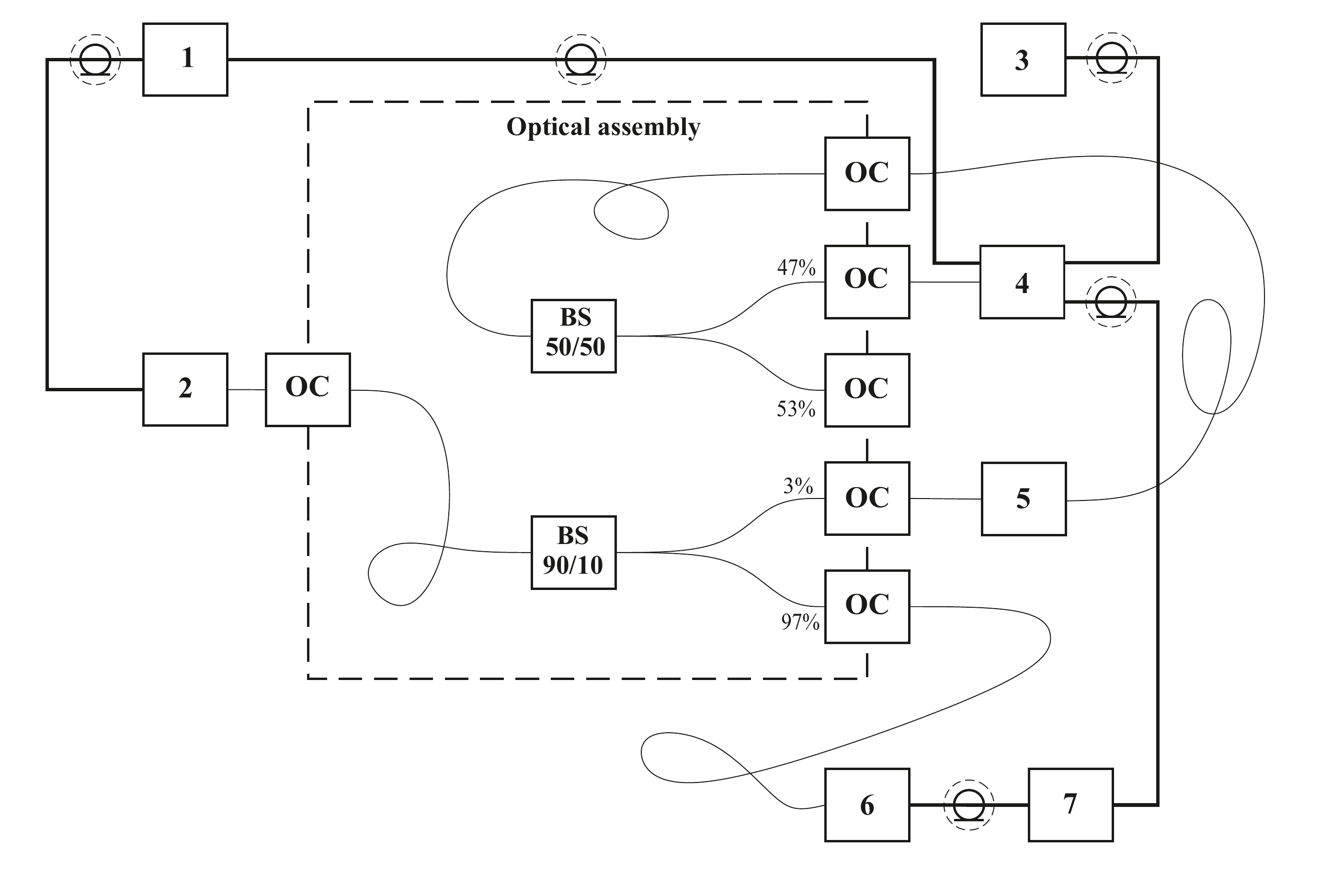}
\caption{Functional scheme of SPD’s measurement setup: 1 -- synchronization system, 2 –- laser pulses source 1550 nm wavelength, 3 -- frequency counter Keysight 53230A, 4 –- tested SPD, 5 -– a system of controlled optical attenuators with output power control, 6 –- multiphoton detector NewFocus Model \#1014, 7 -– oscilloscope Lecroy WaveMaster 830Zi-B-R, OA –- optical adapter, BS –- optical beam splitter.}
\label{fig:meas_sch}
\end{figure}

\qquad We have used the setup shown in Figure  \ref{fig:meas_sch} to measure the SPD’s parameters. This setup includes a synchronization system (1), laser source (2), an optical assembly with optical adapters and beam splitters, a frequency counter (3), a system of controlled optical attenuators with output power control (5), multiphoton detector (6) and oscilloscope (7).

Laser source emits optical pulses at $1550$ nm wavelength. These pulses are generated in the same frequency grid as the gating signal. In the experimental data we present below the repetition rate of laser pulses is $100$ kHz. The delay between gating signal and optical pulses is controlled with the synchronization system. Optical pulses arrive at the input of 90/10 beam splitter, then they are split into two beams with different intensities. A beam with a higher intensity arrives at the input of a wideband optical detector, in which electrical output is connected to an oscilloscope. This detector is used to control the shape, duration, and repetition rate of optical pulses incidented on the SPD  (see Fig. \ref{fig:gate1}b)). 

A beam with a lower intensity arrives at the two series-connected controlled attenuators system input. Due to the possibility of measuring the power of light radiation after the first attenuator and a fixed attenuation coefficient of the second attenuator, it is possible to adjust and maintain the power of laser pulses at the output of about $0.2$ photons per pulse. Then optical pulses arrive at the 50/50 beam splitter, and after it, two detectors can be connected to the measuring system simultaneously. There is 0.1 photon per pulse arrives at the input of each detector after beam splitter.

The signal, attenuated to a level of  $0.1$ photon per pulse, arrives at the tested SPD, where the photon is detected. The number of one-photon states significantly exceeds the number of two-photon states at given energy of light radiation \cite{Gisin}, which ensures the maximum approximation of the results of stand measurements of the SPD performance to the real SPD performance in the QKD setup. The output signal from the detector is incidents simultaneously to a frequency counter and an oscilloscope using an electric power divider. $QE$ and $DCR$ are determined by indications of the frequency counter. The oscilloscope displays a time histogram of triggers, which determines the dead time and the afterpulsing probability.

\section{Results}
\begin{figure}[h]
\begin{minipage}[h]{0.47\linewidth}
\center{\includegraphics[width=1\linewidth]{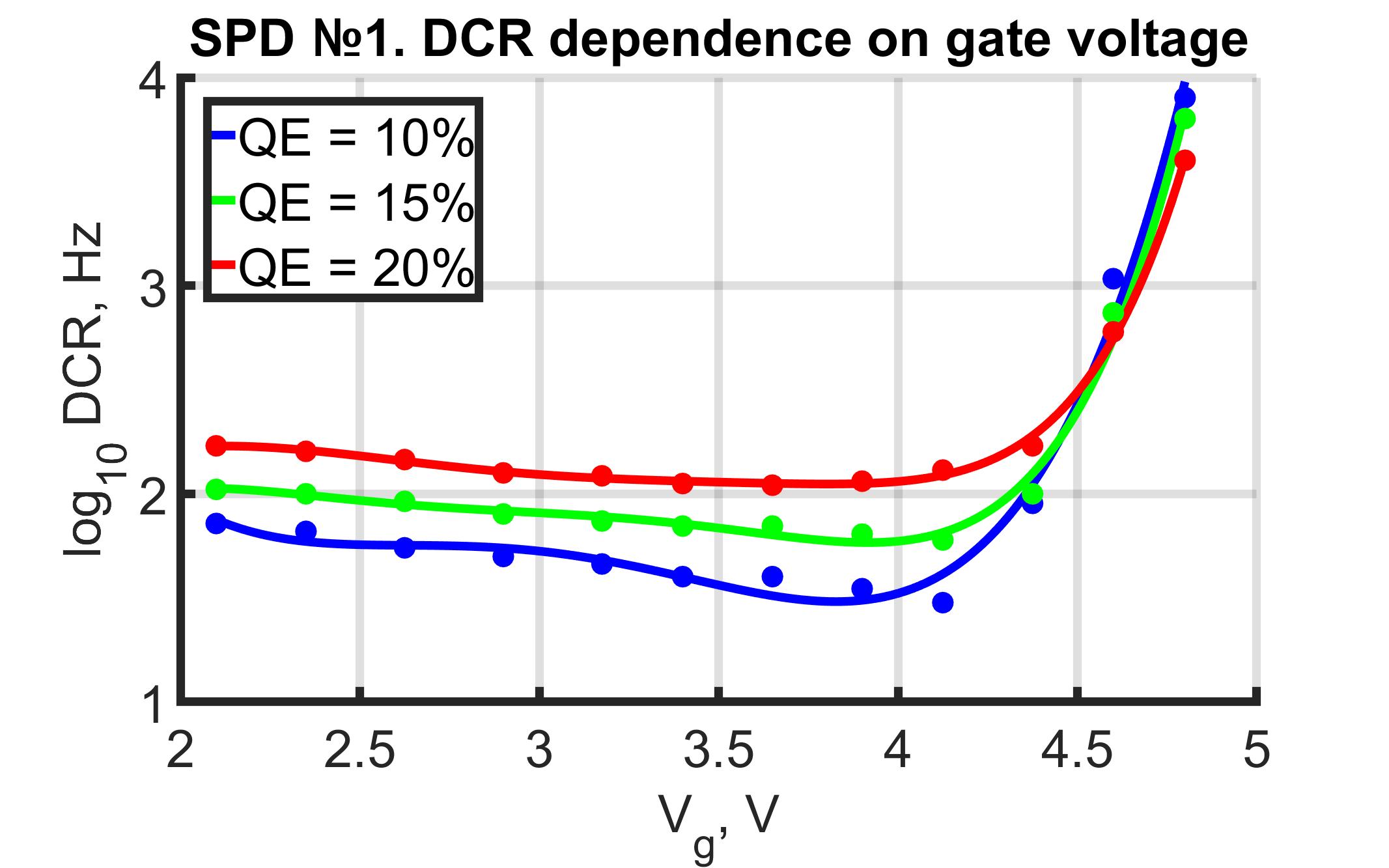}} a) \\
\end{minipage}
\hfill
\begin{minipage}[h]{0.47\linewidth}
\center{\includegraphics[width=1\linewidth]{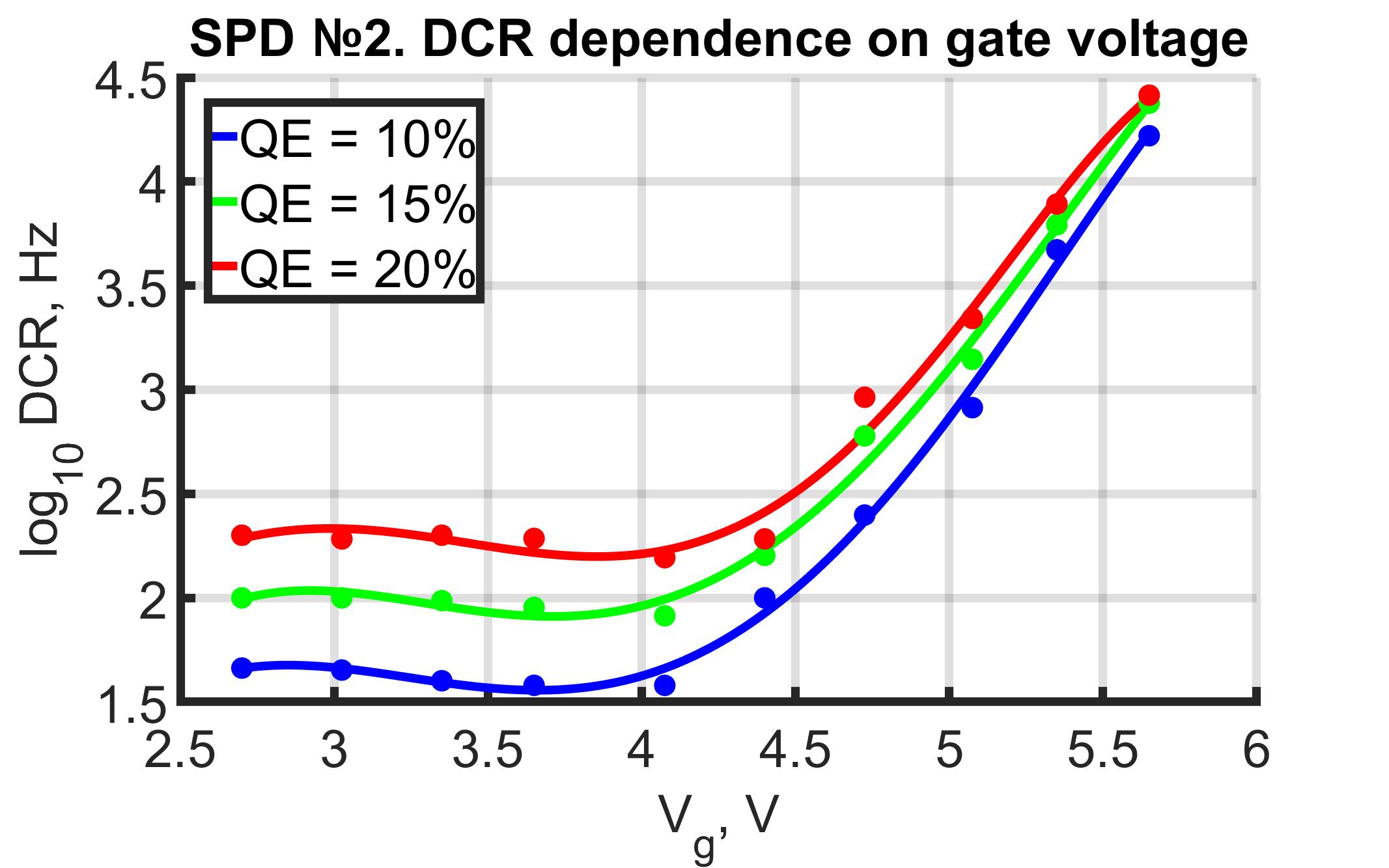}} \\b)
\end{minipage}

\begin{center}
\begin{minipage}[h]{0.47\linewidth}
\center{\includegraphics[width=1\linewidth]{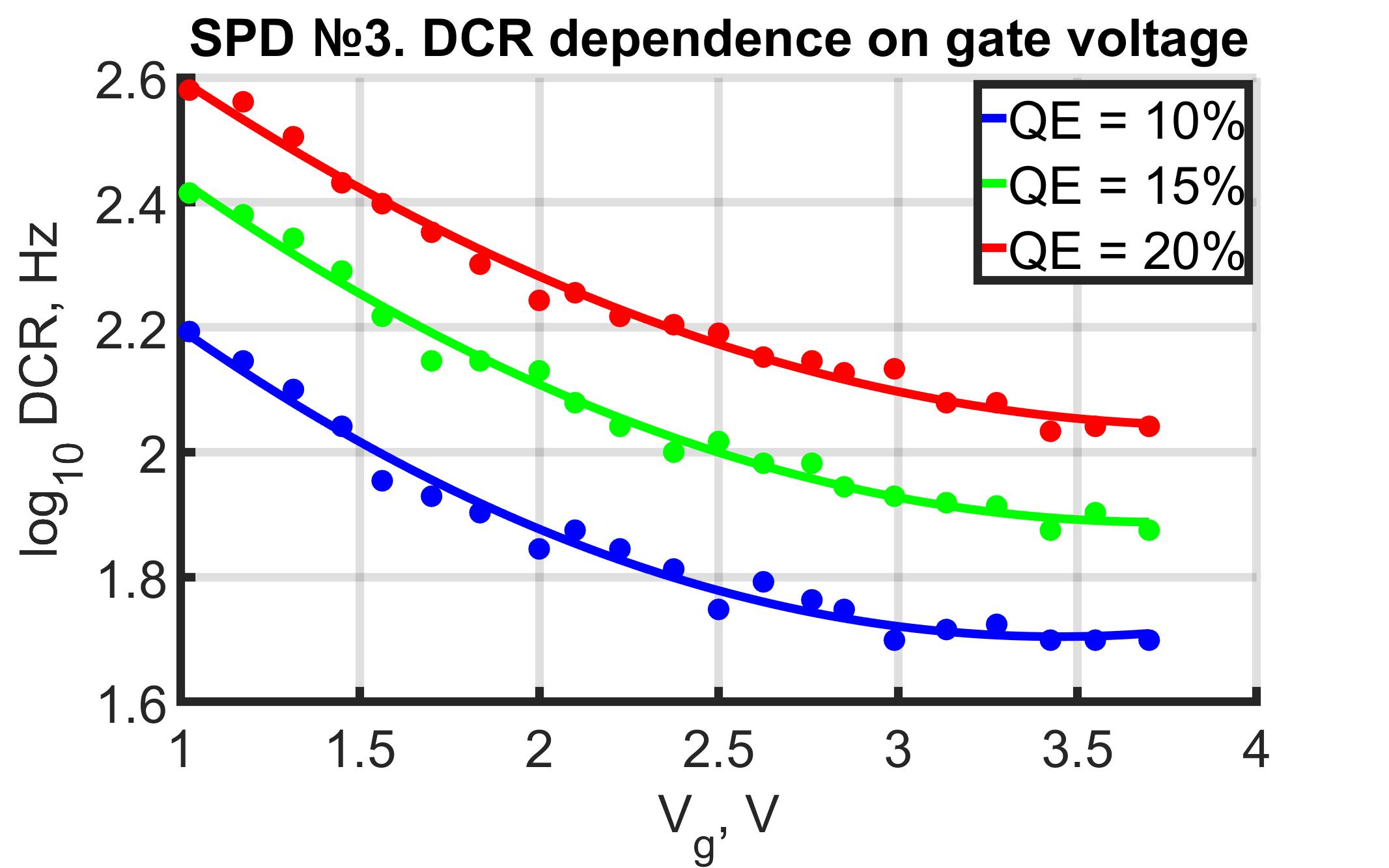}} c) \\
\end{minipage}
\end{center}

\caption{The $DCR$ dependence on the gate amplitude $V_g$ at different values of $QE$ for SPDs with different SPADs: a) SPD №1; b) SPD №2; c) SPD №3.}
\label{fig:dcr1}
\end{figure}

\qquad In this work, we determine the effective mode of SPD’s operation as ensuring the minimum $DCR$ value at a given $QE$ value. To identify the effective operating mode of the SPD, the $DCR$ dependence on the gate amplitude has been measured at constant values of the quantum efficiency, such as $10\%$, $15\%$, $20\%$, and at a maintained temperature on the SPAD equal to $228$ K. The measurement results for the three detectors are shown in Figure \ref{fig:dcr1}.

For SPD №1 and SPD №3 with an increase of gating amplitude $DCR$ level decreases (Fig. \ref{fig:dcr1} a), c)) due to the decrease of Geiger time $t_g$ (see Fig. \ref{fig:gate1}). For SPD №2, the $DCR$ level does not decrease as much as for the other tested SPDs (Fig. \ref{fig:dcr1} b)). We can see $DCR$ decrease by more than two times with an increase of gating voltage from $2.1$ V to $4.1$ V at $QE = 10\%$ for SPD №1. $DCR$ decreases more than three times with an increase of gating signal amplitude from $1$ V to $3.7$ V at $QE = 10\%$ for SPD №3.

There is a sharp increase in the $DCR$ value for SPD №1 and SPD №2 starting from a certain value of the gating signal amplitude. This effect is conditioned by the physical processes, occurring in the SPAD of the corresponding SPD. We suggest that these processes may occur because of the reasons presented below.

The size of the depletion region of SPAD constantly oscillates because the bias voltage on the SPAD is sinusoidal and the redistributed charges have some inertia. These processes determine the electrical capacity of the heterostructure. We consider that with high-amplitude voltage oscillations, a free charge carrier remains near the absorption region. This charge subsequently leads to the generation of an avalanche process and contributes to the $DCR$.

We also consider two other hypotheses of the origin of the effect of a sharp increase in $DCR$. It should be noted that these hypotheses have been experimentally falsified. However, the studies we have carried out make it possible to more fully understand the processes occurring in the SPD and to build regularities between the main parameters. It can be used to optimize the SPD’s operation.

\subsection{The increase in the probability of tunneling generation hypothesis}
\qquad The first hypothesis for a sharp increase in $DCR$ is an increase in the probability of tunnel charge generation, which may be due to an extremely sharp increase in the excess bias voltage  $V_{ex}$. 

\begin{figure}[h]
\begin{minipage}[h]{0.47\linewidth}
\center{\includegraphics[width=1\linewidth]{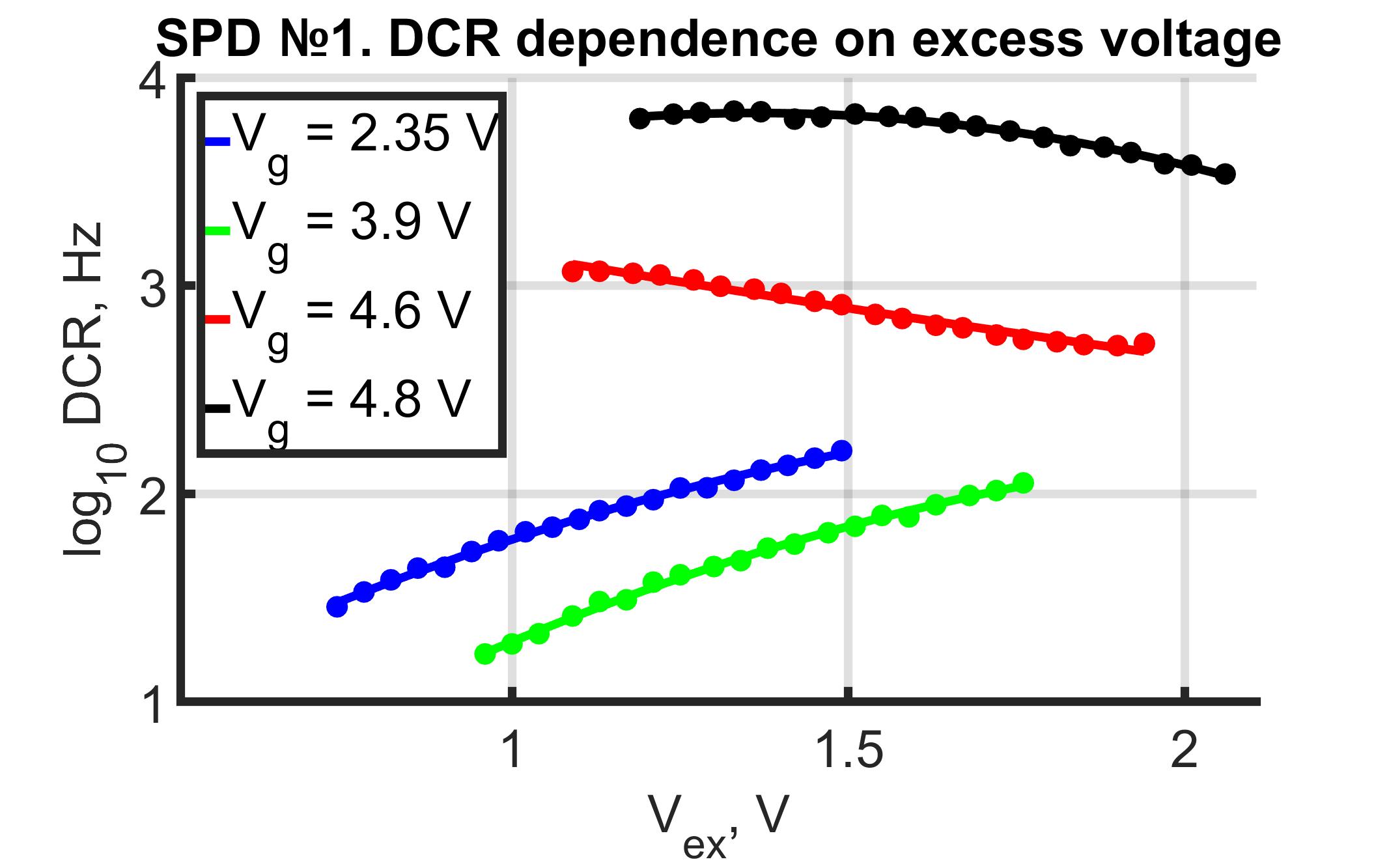}} a) \\
\end{minipage}
\hfill
\begin{minipage}[h]{0.47\linewidth}
\center{\includegraphics[width=1\linewidth]{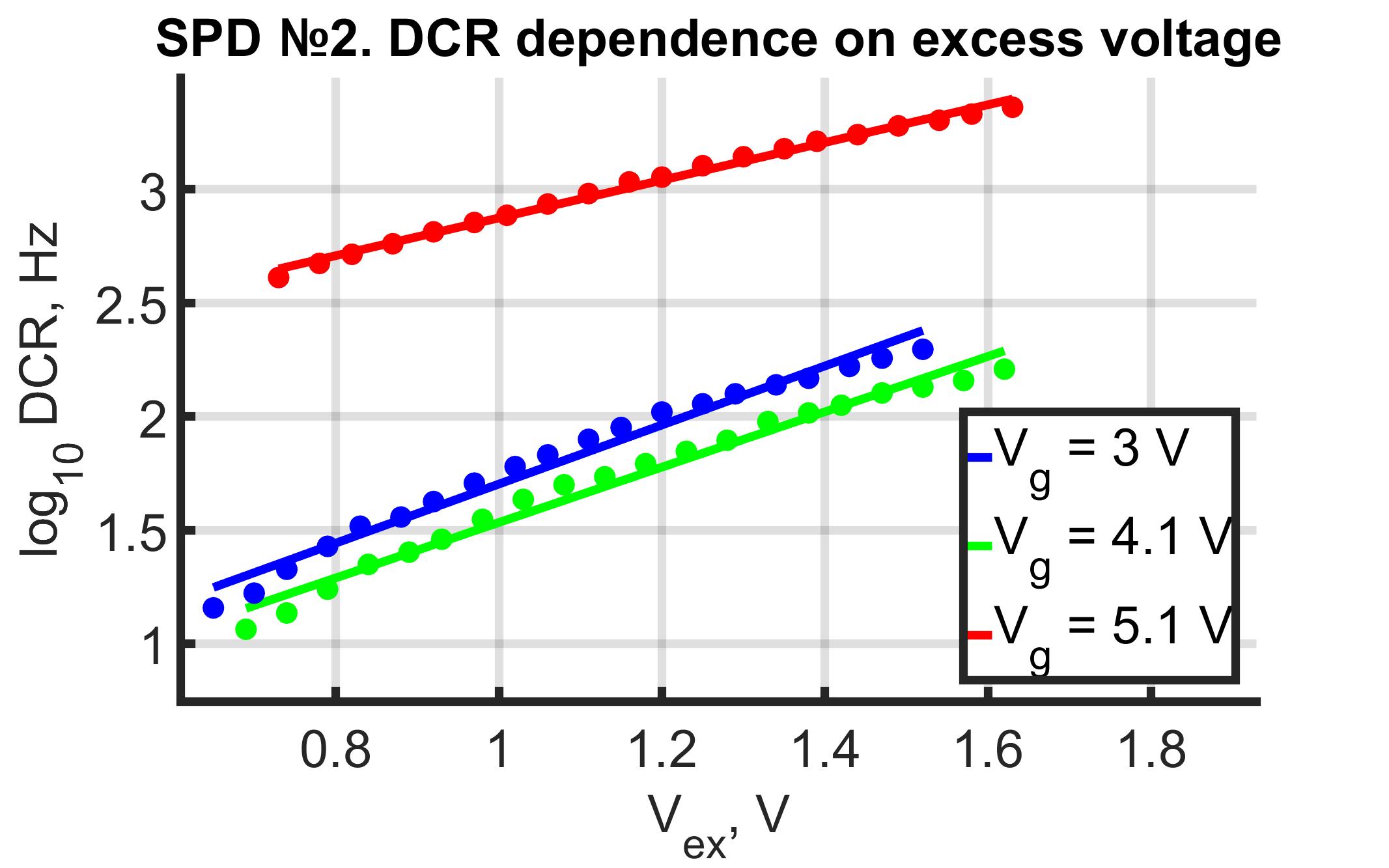}} \\b)
\end{minipage}

\begin{center}
\begin{minipage}[h]{0.47\linewidth}
\center{\includegraphics[width=1\linewidth]{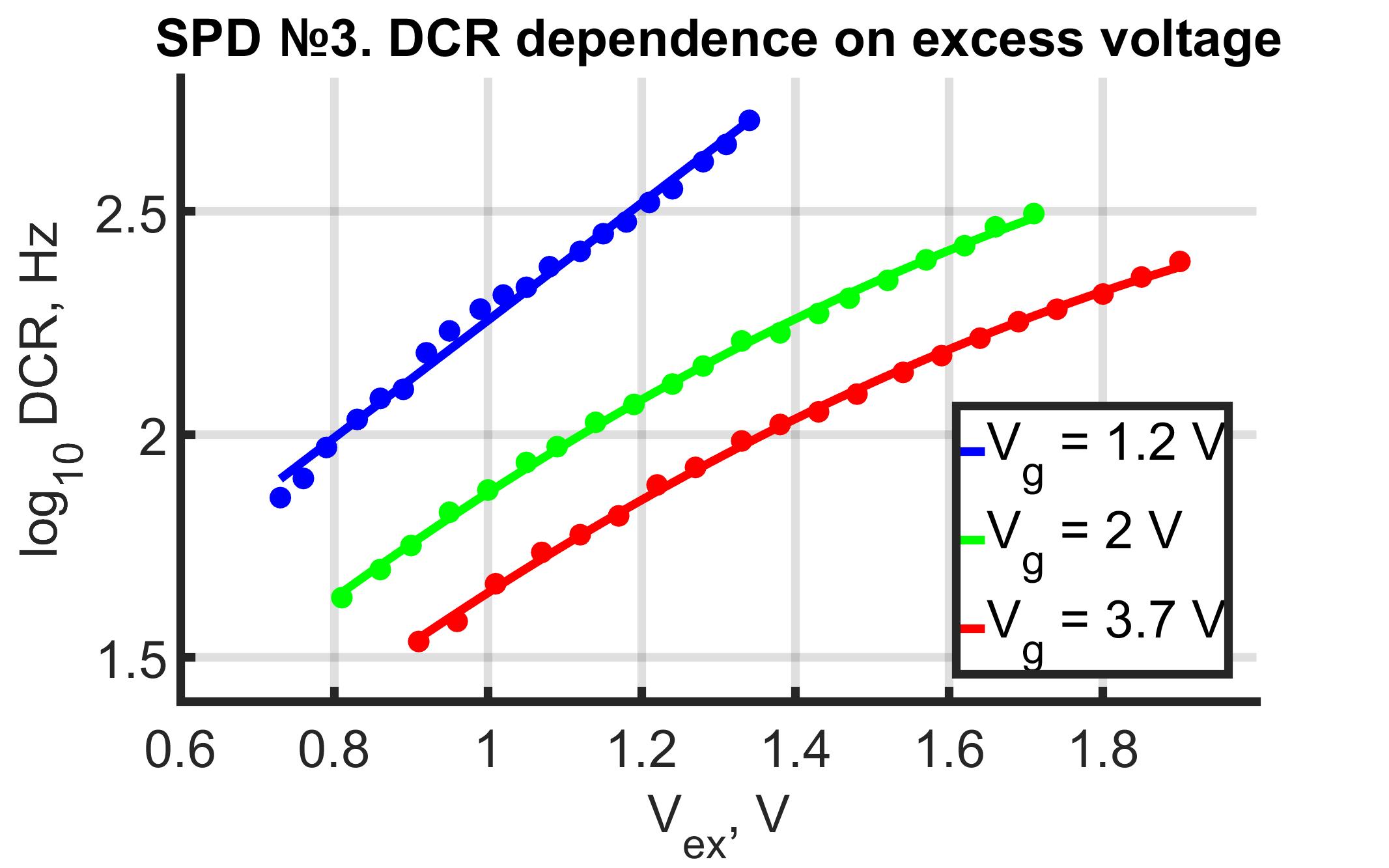}} c) \\
\end{minipage}
\end{center}

\caption{The $DCR$ dependence on excess voltage  $V_{ex}$ at different values of the gating signal amplitude $V_g$ for SPDs with different SPADs: a) SPD №1; b) SPB №2; c) SPD №3. }
\label{fig:dcr2}
\end{figure}

The dependence of $DCR$ on excess voltage $V_{ex}$ at different values of the gating signal amplitude is shown in Figure \ref{fig:dcr2}. For SPD №1 we can see that with the same value of excess voltage $V_{ex}$, $DCR$ with gating amplitude values $V_g = 4.6$ V and $V_g = 4.8$ V significantly exceeds $DCR$ with gating amplitude values $V_g = 2.35$ V and $3.9$ V (Fig. \ref{fig:dcr2}a). A similar situation is observed for SPD №2: at the same value of excess voltage $V_{ex}$, $DCR$ with gating amplitude value $V_g = 5.1$ V significantly exceeds $DCR$ with gating amplitude values $V_g = 3$ V and $V_g = 4.1$ V (Fig. \ref{fig:dcr2}b).

An increase in the $DCR$ value with increasing  $V_{ex}$ and a constant gating signal amplitude is shown in the graphs in Figure  \ref{fig:dcr2}. It is mainly due to an increase in Geiger time. A change in the excess voltage  $V_{ex}$ within $[0, \ 2]$ V about the breakdown voltage  $V_{br} \approx 70$ V does not have such a strong effect on the probability of tunneling generation.

It should be noted that the graph for SPD №1 with gate amplitude values  $V_g = 4.6$ V and $V_g = 4.8$ V has an effect associated with a decrease in the $DCR$ value with an increase in excess voltage $V_{ex}$.  We find it difficult to put forward an assumption about the cause of this effect. We will consider it in future works.
 
\subsection{The decrease in the photon detection probability hypothesis}

\qquad The second hypothesis for a sharp increase in $DCR$ is a sharp decrease in the $QE$ value with an increase in the gate amplitude and a constant excess voltage  $V_{ex}$.

\begin{figure}[h]
\begin{minipage}[h]{0.47\linewidth}
\center{\includegraphics[width=1\linewidth]{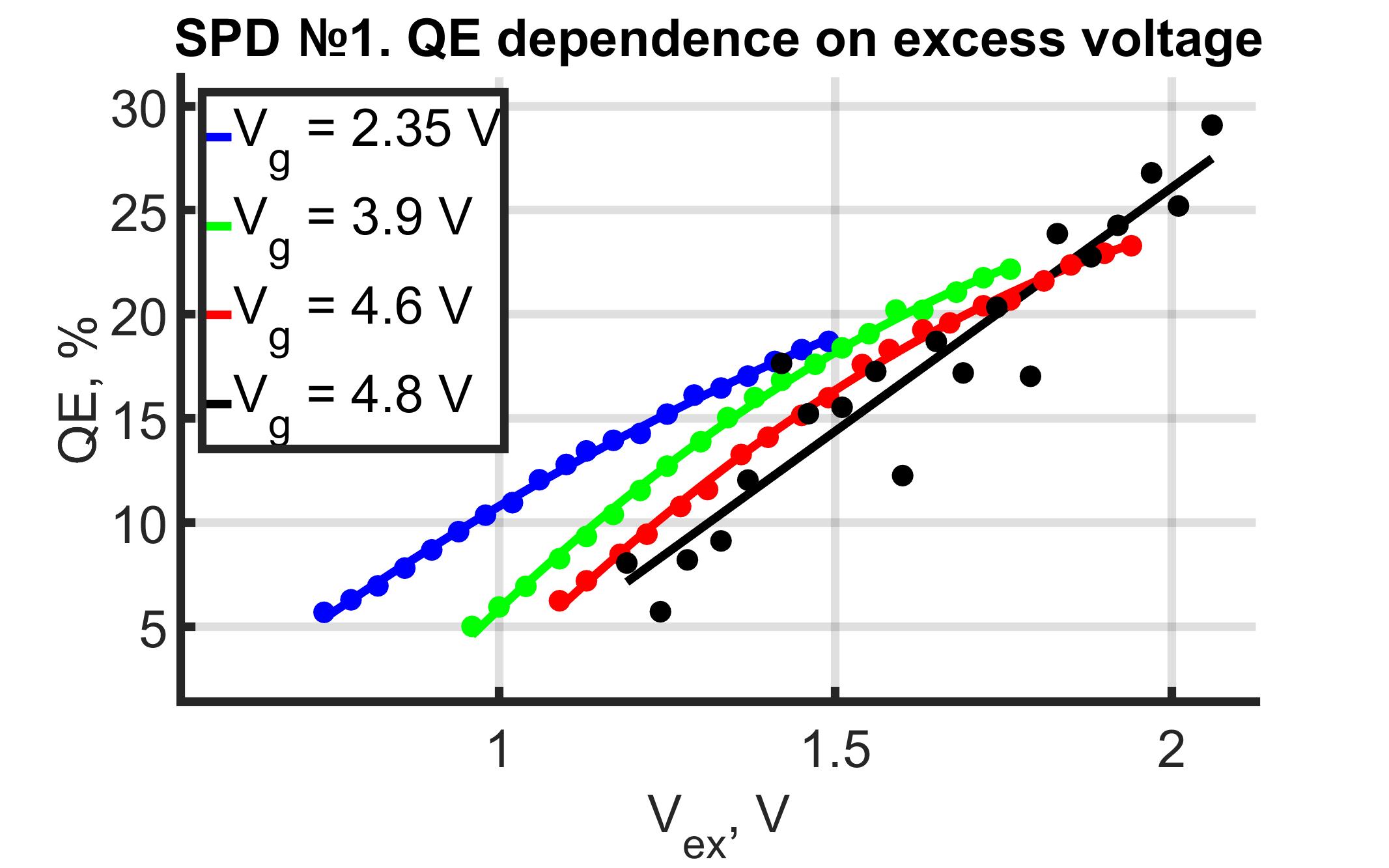}} a) \\
\end{minipage}
\hfill
\begin{minipage}[h]{0.47\linewidth}
\center{\includegraphics[width=1\linewidth]{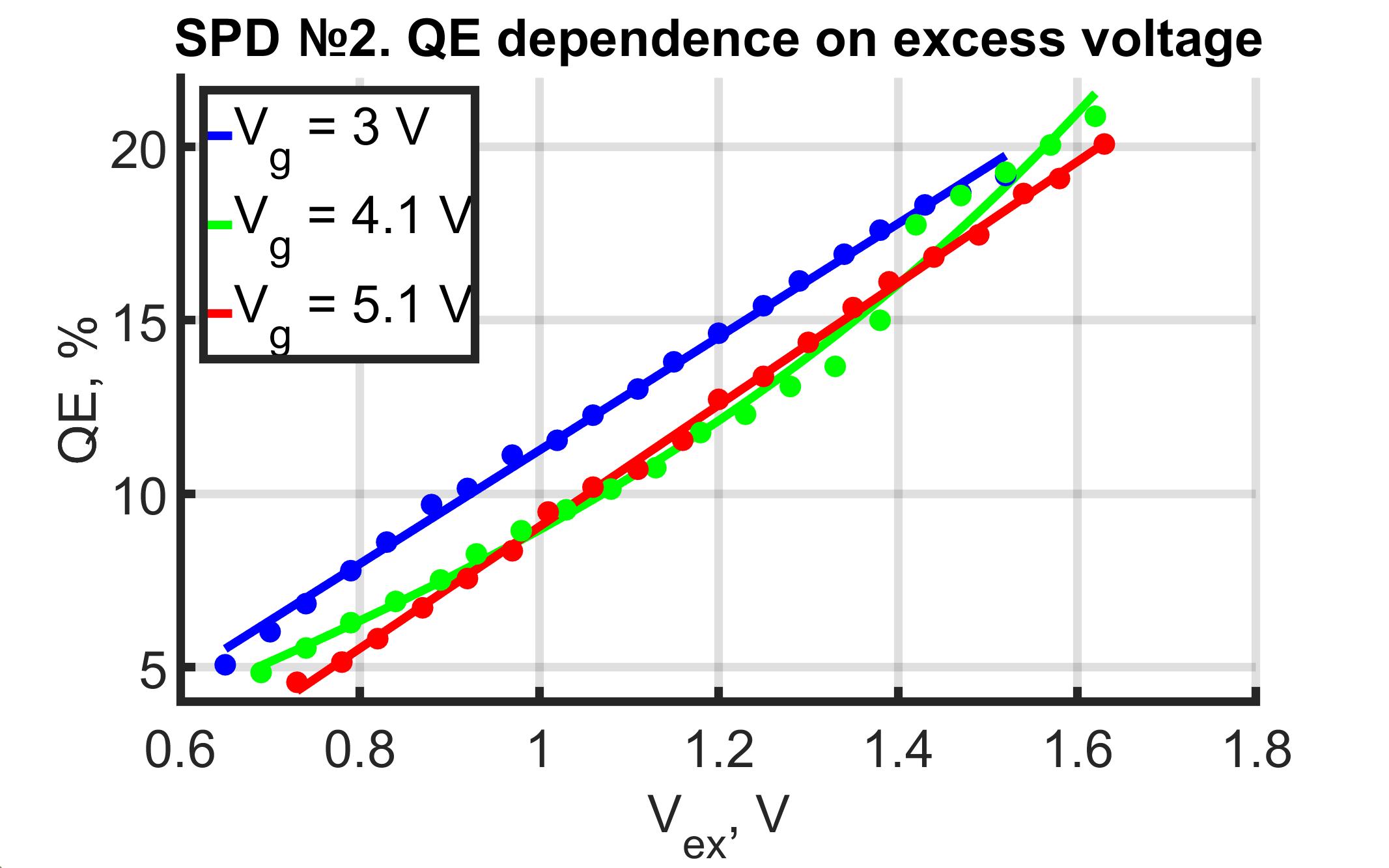}} \\b)
\end{minipage}

\vfill
\begin{center}
\begin{minipage}[h]{0.47\linewidth}
\center{\includegraphics[width=1\linewidth]{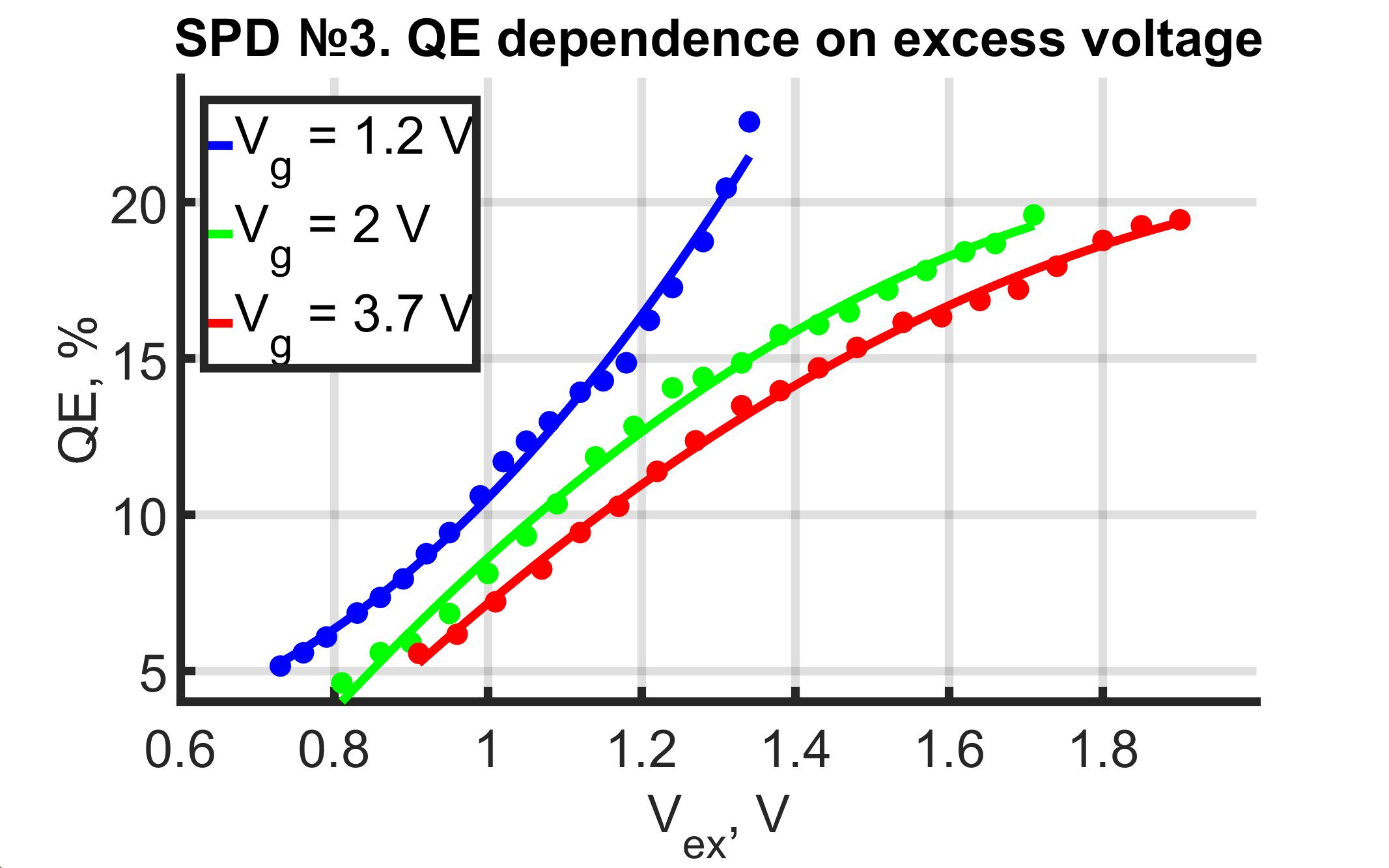}} c) \\
\end{minipage}
\end{center}

\caption{The $QE$ dependence on excess voltage $V_{ex}$ at different values of the gating signal amplitude $V_g$ for SPDs with different SPADs:  a) SPD №1; b) SPB №2; c) SPD №3.}
\label{fig:qe1}
\end{figure}

We show the dependence of $QE$ on the excess voltage $V_{ex}$ at different values of the gating amplitude in Figure \ref{fig:qe1}. With an increase in $V_{ex}$ and a constant gating amplitude, an increase in $QE$ is observed in all graphs due to an increase in the probability of transporting the primary electric charge from the absorption region to the multiplication region. These processes also cause an increase in the probability of an avalanche process.

With an increase in the gating amplitude and a constant value of  $V_{ex}$, a decrease in the value of $QE$ due to a decrease in the Geiger mode time $t_g$ is observed. Firstly, this effect is due to a decrease in the probability of a photon arriving in a time window $t_g$ due to the presence of a timing jitter of the photon emission time. Even in the case of a photon arriving at SPAD at the specified time window, the avalanche generation may not occur, since the primary electric charge may appear in the multiplication region at the moment when the bias voltage on SPAD is less than the breakdown voltage. Secondly, the most favorable time interval for detection, characterized by high values of the excess voltage  $V_{ex}$, is reduced.  

We discovered that this hypothesis is not the reason for the observed effect of a sharp increase in $DCR$. As we can see in the Figure \ref{fig:qe1}a,  the decrease in $QE$ with a constant value of $V_{ex}$ and an increase in the gating signal amplitude occurs by approximately equal values for SPD №1. It is necessary to increase $V_{ex}$ also by approximately equal values to maintain $QE$ at the same level. Thus, with gate amplitude values $V_g = \{4.8, 4.6, 2.35, 3.9\}$ V, there are similar trends in the change in the probability of photon detection.

Similar statements can be made for SPD №2. The described effect was also observed on this detector. However, for the gate amplitude value $V_g = 4.1$ V, at certain values of $V_{ex}$, the probability of photon detection is higher than at $V_g = 5.1$ V (Fig. \ref{fig:qe1}b).

We also can see all the described tendencies of $QE$ change for the SPD №3 (Fig. \ref{fig:qe1}c). However, there is one more interesting effect. For $V_g = 1.2$ V, $QE$ grows much faster than for $V_g = 2$ V and $V_g = 3.7$ V.

Three SPDs with identical structures and different SPAD diodes were measured. We can conclude that define a universal empirical formula that could relate the main operational parameters of SPDs with the main parameters of the control voltage, such as $QE$ and $DCR$ with $V_{b1}$ and $V_g$, is not possible at this stage of research. This is due to errors that arise during the production of the SPAD heterostructure. It is observed even in one batch of devices and makes it difficult to formalize the dependence for the main SPAD parameters, as shown, for example, in \cite{compstudy}. 

Based on the results obtained, it is possible to formulate general recommendations for setting up an SPD with a similar control electrical circuit. To decrease the $DCR$, the gating signal amplitude should be increased, starting from the lowest possible value, with the $QE$ unchanged, until the moment when a strong increase of $DCR$ begins.

\section{Conclusion}
\qquad In this work, we have studied the influence of the gating signal parameters on the main operational parameters of the SPD: $QE$ and $DCR$. The first result of the work is the proposed universal recommendations for increasing the performance of the SPD with a sinusoidal gate signal, passive quenching, and active or passive reset. Recommendations are based on the established relationship of $DCR$ decrease with increasing gating signal amplitude, starting from the lowest possible value, up to a certain boundary value. However, when reaching a certain amplitude of the gate, a sharp increase in the $DCR$ value begins. Using the described dependence has made it possible to reduce the $DCR$ by up to three times. The presented method for reducing the noise characteristics is based on reducing the Geiger mode time $t_g$. The results obtained make it possible to optimize the operational parameters of the sine-wave gated SPD. It improves the operational parameters of the whole QKD device.

The second result is the observation of the effect of a sharp increase in the $DCR$ value with increasing gating signal amplitude and constant $QE$. We have assumed the possible interpretation of this effect. A free charge carrier remains near the absorption zone with high-amplitude oscillations, which subsequently leads to the generation of an avalanche process. We perceive such avalanches as noise. It is necessary to exclude the hypothesis of a sharp increase in $DCR$ because of tunneling charge generation, or a sharp decrease in $QE$ with an increase in the gating signal amplitude and a constant excess voltage $V_{ex}$. It has been shown experimentally. A more detailed study of the nature of this effect based on numerical modeling will be presented in a separate work.

The third result is confirmation of the conclusion of the article \cite{compstudy}. It consists in the difficulty of drawing up a general empirical dependence for individual operational parameters of the SPD, that’s why we need to study each device with its mathematical model. However, in \cite{compstudy} only the afterpulse effect was considered. In this work, we have concluded that similar difficulties arise when trying to empirically relate the main operational parameters of the SPD with the main parameters of the control voltage, namely $QE$ and $DCR$ with $V_{b1}$ and $V_g$.

\bibliography{iter8.bbl}
\end{document}